\documentclass[journal]{IEEEtran}

\ifCLASSINFOpdf
\else
\fi

\usepackage{graphicx}
\usepackage{url}
\usepackage{amsmath}
\usepackage{amssymb}
\usepackage{bm}
\usepackage{subfigure}
\usepackage{algorithm}
\usepackage{multirow}
\usepackage{url}
\usepackage{balance}
\usepackage{epstopdf}
\usepackage{txfonts}
\usepackage{color}
\usepackage{comment}
\usepackage{multirow}
\usepackage{hhline}
\usepackage{epstopdf}

\begin{document}
	
\title{Caching-Aided Collaborative D2D Operation for Predictive Data Dissemination in Industrial IoT}

\author{Antonino Orsino, Roman Kovalchukov, Andrey Samuylov, Dmitri Moltchanov,\\Sergey Andreev, Yevgeni Koucheryavy, and Mikko Valkama
\thanks{\textcopyright \ 2018 IEEE. Personal use of this material is permitted. Permission from IEEE
	must be obtained for all other uses, in any current or future media, including
	reprinting/republishing this material for advertising or promotional purposes,
	creating new collective works, for resale or redistribution to servers or lists, or
	reuse of any copyrighted component of this work in other works. 
	
	The authors are with the Laboratory of Electronics and Communications Engineering, Tampere University of Technology,~Finland.}
}

\maketitle

\begin{abstract}
Industrial automation deployments constitute challenging environments where moving IoT machines may produce high-definition video and other heavy sensor data during surveying and inspection operations. Transporting massive contents to the edge network infrastructure and then eventually to the remote human operator requires reliable and high-rate radio links supported by intelligent data caching and delivery mechanisms. In this work, we address the challenges of contents dissemination in characteristic factory automation scenarios by proposing to engage moving industrial machines as device-to-device (D2D) caching helpers. With the goal to improve reliability of high-rate millimeter-wave (mmWave) data connections, we introduce the alternative contents dissemination modes and then construct a novel mobility-aware methodology that helps develop \textit{predictive} mode selection strategies based on the anticipated radio link conditions. We also conduct a thorough system-level evaluation of representative data dissemination strategies to confirm the benefits of predictive solutions that employ D2D-enabled collaborative caching at the wireless edge to lower contents delivery latency and improve data acquisition reliability.
\vspace{-0.4cm}
\end{abstract}



\section{Collaborative Edge Caching for Industrial IoT}\label{sect:intro}

In recent years, there has been considerable progress in data caching and delivery techniques, especially for mobile applications and services. Conventionally, mobile contents are cached in the intermediate servers to reduce transmitted traffic volumes and avoid duplicate downloads of selected popular data~\cite{6736753}, such as viral video files with high revisit rates. This approach helps alleviate extensive response times to fetch the more demanded data at the mobile device as well as improves spectral efficiency and reduces energy consumption~\cite{7565183}. Subject to appropriate delivery and placement strategies, popular contents may be reused when accessed asynchronously (i.e., at different times) by a multitude of data consumers.

Historically, caching has been most explored to facilitate delivery of non-real-time contents, such as video-on-demand. The many strategies proposed to date primarily catered for the optimum balance between the transmission costs (in terms of e.g., wireless bandwidth required for data delivery) and the storage costs (as memory capacity becomes more affordable today). However, the proliferation of advanced Internet of Things (IoT) applications is driving the storage and computing resources toward further dispersion~\cite{7807196}. This means that storage nodes are deployed at the network edge while the classical communication---storage trade-off shifts to employing \textit{edge caching} for increased contents availability and improved reliability of data delivery.

Placed in closer proximity to mobile IoT devices, additional storage capacity improves end-to-end data transfer latency, especially when the human operator is also located on-site e.g., in a nearby control cabin. In industrial applications, lower latency translates into better responsiveness while producing backup copies of critical contents improves scalability, availability, and reliability of mission-aware applications. This is particularly important to support \textit{surveying and inspection} operations where a fleet of industrial vehicles, robots, or drones observes factory facilities, mines, harbors, and construction sites in remote and high-risk locations. These intelligent moving machines may produce high-definition video streams and other heavy sensor data~\cite{Tor15}, which need to be transferred to the human operator in a delay-sensitive manner and potentially on-demand. The latter requires reliable and high-quality radio communication links between the industrial machines and the control station.

\begin{figure}[!ht]
\centering
\includegraphics[width=1.0\columnwidth]{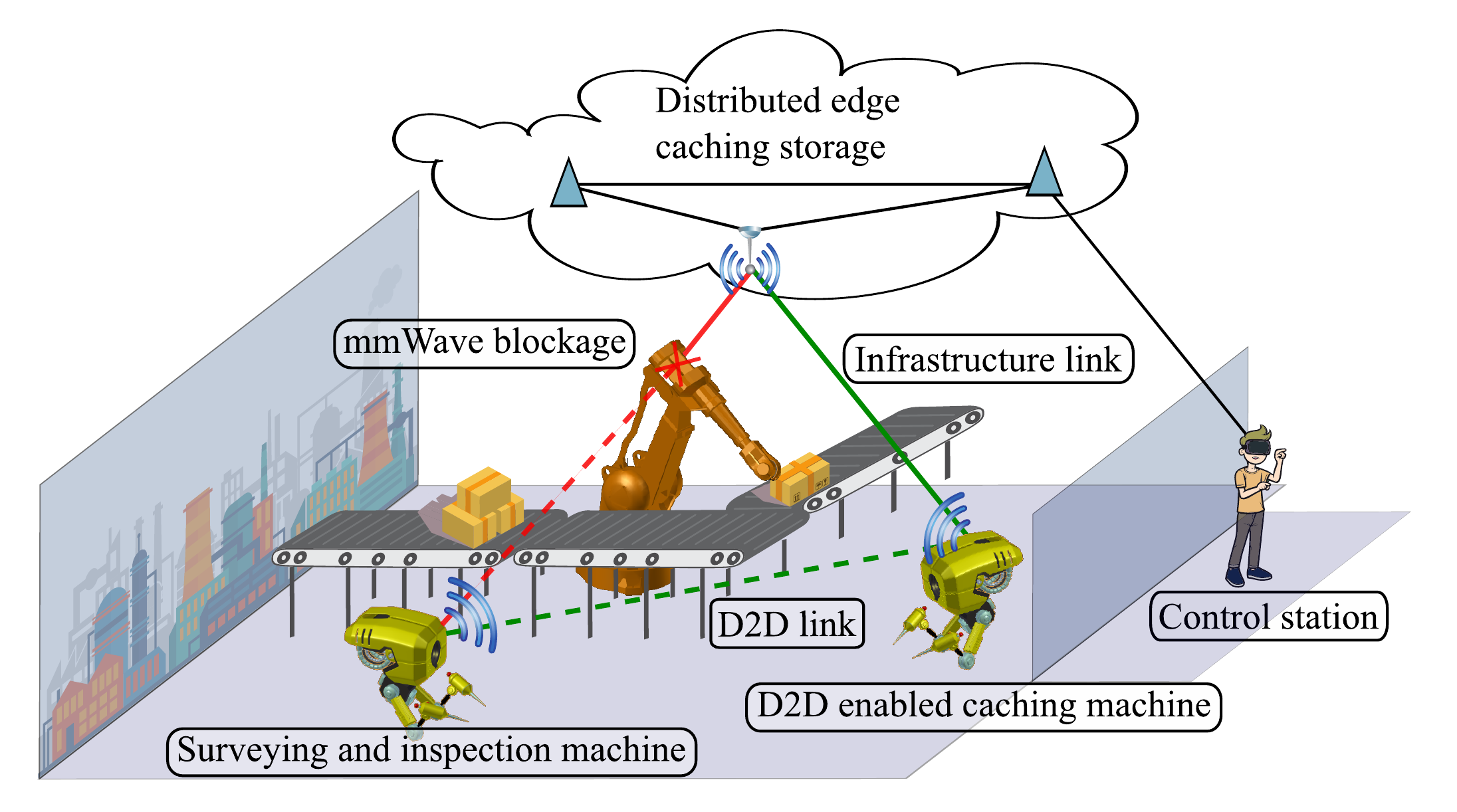}
\caption{Industrial IoT with collaborative in-device caching.}
\label{fig:concept}
\end{figure}

However, industrial IoT environments are typically large and challenging locations where provisioning for seamless high-rate wireless connectivity is cumbersome. Fifth-generation (5G) mobile technology is a radical innovation that attempts to improve over this situation by outlining a New Radio (NR) solution operating in millimeter-wave (mmWave) frequencies. While offering much higher bandwidths, mmWave links are susceptible to many adverse effects, such as dynamic blockage by smaller obstacles, which need to be taken into account comprehensively (see Fig.~\ref{fig:concept}). As a result, mission-critical data of future industrial IoT applications will benefit from high-rate connections, but contents availability, delivery reliability, and acquisition latency may still be disadvantaged without intelligent edge caching solutions.

Fortunately, today's IoT machines are more advanced devices that can afford to cache popular contents locally by facilitating delay-aware multimedia delivery in surveying and inspection operations. Accordingly, industrial machines moving at various speeds may exchange and retrieve from each other partial video segments as requested by the human operator(s) currently. Such intelligent precaching at mobile devices equipped with 5G-grade device-to-device (D2D) communication capabilities constitutes a flexible and cost-efficient deployment option that augments static 5G mmWave layouts to reduce harmful outages. Very different from caching in the core network, \textit{collaborative} D2D-enabled caching at the wireless edge has the potential to better accommodate immediate service quality requirements and traffic demand profiles, but requires a careful research perspective. Systematically delivering it becomes the primary focus of this work and the next section discusses the key benefiting applications.


\section{D2D-Enabled Caching in Factory Automation}\label{sect:app}

Industrial automation emerges as an important economy sector where various products are assembled, tested, or packaged in several well-defined production stages (automotive, general consumer electronics, goods production, etc.). For factory automation, surveying and inspection operations require on-time delivery of heavy traffic with high levels of reliability to avoid unwanted intermissions in the manufacturing process. Typically, every production stage involves interactions with a multitude of sensors and actuators controlled centrally (e.g., with a programmable logical controller). In the past, many of these connections used to be wired, which made them endure the stress of repeating movements in harsh industrial environments. For this reason, the in-factory sensors and actuators are now increasingly accommodated with wireless technology to improve the flexibility and augment the availability of their produced data.

As today's factories are becoming increasingly complex and wireless, they require frequent surveying and inspection operations that may employ fleets of small-sized industrial vehicles or robots that observe the factory facilities and report on their status to the human operator(s) located inside the control station. These moving machines can perform autonomously (or be controlled remotely via AR/VR equipment) and produce bandwidth-hungry video streams together with other heavy sensor data~\cite{Tor15}. Consequently, high-rate mmWave communication links are needed to deliver such delay-sensitive traffic from the remote machines to the on-site operator(s). Further, the moving robots can also be arranged into virtual teams that may exchange and retrieve heavy maintenance data collaboratively (see Fig.~\ref{fig:concept}) subject to appropriate incentives. However, adequate reliability and controlled latency are essential to ensure that such mission-aware operations run without harmful interruptions~\cite{Ericsson_demo}.

To support timely dissemination of collaborative contents at the edge, in-device caching is envisioned as an effective technology solution to satisfy the stringent requirements of industrial surveying and inspection applications. The benefits that collaborative edge caching brings to the factory automation contexts are significant. In this work, we address a characteristic scenario where moving industrial machines produce and exchange bandwidth-demanding (e.g., multimedia) contents collaboratively. They also employ in-device caching and D2D communication capabilities to compensate for the potential intermittency of mmWave connections, and thus lower the response times as well as improve the operational reliability. D2D-enabled caching at the edge can also alleviate backhaul requirements and enjoy higher degrees of spatial reuse while facilitating local distribution of delay-sensitive traffic. The following section explains the nature of reliability-related concerns in the considered use cases.


\section{Caching-Aided Reliability over mmWave Connections}

It is becoming widely understood that advanced industrial IoT applications will require mmWave connections, particularly in surveying and inspection operations that produce high-rate video streams. Because of the potential latency constraints, such services may not use compression of multimedia data, which adds to the link capacity requirements that can only be satisfied with abundant mmWave bandwidths. However, the link-layer performance of 5G mmWave systems is drastically different from that at microwave frequencies. Even though the utilization of higher carrier frequencies and larger bandwidths promises to enable much better transmission rates, mmWave bands introduce their specific challenges to the respective system design. At these frequencies, the path loss becomes significantly higher, which limits the achievable coverage ranges. This issue can be mitigated by exploiting highly directional antennas at both ends of a communication link.

Another specific feature of mmWave systems, which operate at wavelengths of under a centimeter, is that not only the larger objects -- such as (parts of) buildings -- affect the radio propagation properties, but also much smaller obstacles become impactful, which may include the elements of the factory floor and conveyor belts, industrial machines themselves, etc. Across the envisioned deployment scenarios of 5G mmWave cellular, 3GPP has identified that human body blockage becomes another major factor in characterizing mmWave propagation~\cite{metis, standard_16}. More recently, a growing number of works addressed the key mmWave propagation phenomena. They confirm that dynamic blockage of the line-of-sight (LoS) link between a mobile device and its serving mmWave station by various objects, such as humans and vehicles, not only causes frequent and abrupt deviations in the amounts of the required radio resources at sub-second timescales, but may also lead to harmful outages~\cite{gapeyenko2017temporal,samuylov2016characterizing}. These adverse effects call for intelligent edge caching mechanisms to improve upon communication reliability over inherently intermittent mmWave links.

To augment reliability of data dissemination in industrial mmWave deployments, a range of potential solutions has been considered, including fallback to other radio-access technologies, bandwidth reservation, and multi-connectivity operation. In integrated multi-radio systems, an active mmWave session can in principle be (partially) offloaded onto LTE or Wi-Fi technology~\cite{mehrpouyan2015hybrid,niu2015survey}. However, the dramatic gap between the data rates of mmWave and microwave radios as well as the bandwidth-hungry nature of industrial IoT applications impose considerable limitations on the use of this method. To complement, multi-connectivity operation in dense mmWave deployments allows to employ several mmWave connections simultaneously to decrease the outage time~\cite{petrov2017dynamic}. Finally, reserving some bandwidth at the mmWave stations exclusively for serving the ongoing data flows may also benefit session continuity.

However, the practical applicability of the above methods is limited to specific environments that feature dense mmWave deployments and may require provisioning of other radio access networks across the service area. This is generally costly and even if achieved the resultant connection reliability may still be inadequate. In industrial automation scenarios, it is more typical that the mobile device is connected to a single serving station and can seldom benefit from multi-connectivity or multi-radio improvements in contrast to outdoor urban deployments. In these situations, collaborative D2D-enabled caching has the potential to augment performance of cell-edge moving devices and thus may become a key mechanism to decisively decrease the outage time in mmWave systems. This is due to significantly higher spatial channel diversity across the neighboring IoT machines. 

Moreover, the highly directional nature of mmWave links may efficiently mitigate radio interference created by several simultaneous transmissions and thus further boost the area capacity~\cite{petrov2017interference}. To make this vision a reality and unlock the potential benefits of reliable edge caching over mmWave, a number of challenges need to be resolved. Because of the bandwidth-hungry nature of the considered factory automation applications, moving industrial machines may have to perform \textit{proactive} caching and intelligent forwarding of crucial data collaboratively~\cite{6871674}. Since the data rates on the mmWave links are substantial, devices need to be equipped with a matching cache size, which is fortunately feasible for the advanced factory automation machines. In what follows, we discuss the available options for caching-aided collaborative data dissemination over mmWave and introduce the distinct operational modes. 


\section{Available Data Dissemination Modes}\label{sec:up_schemes}

This work focuses on the factory automation scenarios where various moving devices, such as industrial vehicles, robots, or drones, are involved collaboratively into surveying and inspection operations. To this aim, they are equipped with advanced capabilities to capture, store, and stream multimedia data (e.g., raw video contents and/or massive sensor readings) that are eventually made available in a delay-sensitive manner or on-demand to the remote human operator(s) located e.g., in a (virtual) control station. The advanced IoT devices are also supplied with the mmWave-based radio technology to reach a base station as well as with the means for D2D connectivity. Accordingly, we differentiate between \textit{three} alternative data dissemination modes depending on the perceived radio link quality and the availability of proximate devices or network infrastructure. 

\begin{figure}[!ht]
\vspace{-0.3cm}
\centering
\includegraphics[width=1.0\columnwidth]{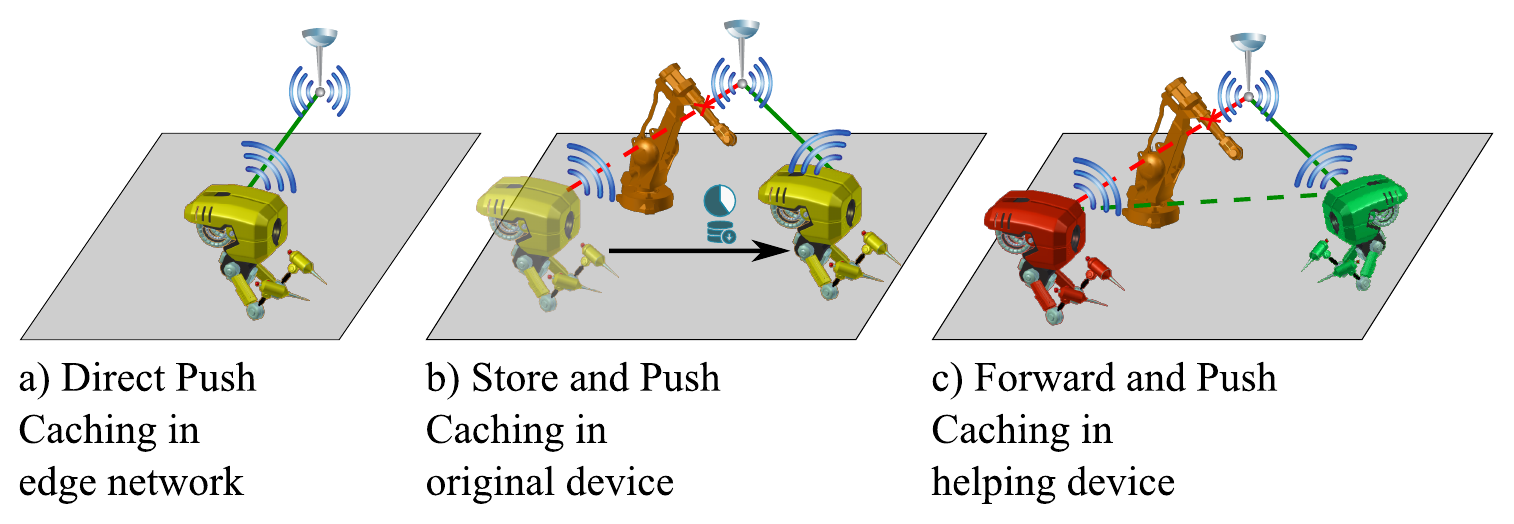}
\caption{Considered data dissemination modes.}
\label{fig:modes}
\vspace{-0.3cm}
\end{figure}

The first mode is named here \textit{Direct Push} and assumes that a moving machine that experiences favorable radio link conditions to its serving mmWave station decides to immediately push its heavy data into the network instead of storing it locally (see Fig.~\ref{fig:modes}(a)). Then, the data are cached in the edge network infrastructure until when they are demanded by any of the potential remote operators (i.e., for control or monitoring purposes). In this situation, the main requirement is the presence of a stable and unobstructed mmWave path, which is not blocked by any obstacle. 

However, as explained in the previous section, it is very typical for a mobile device to experience intermittent mmWave connectivity in non-LoS locations and whenever occluded by another (moving) object. Due to substantial losses in signal strength under non-LoS conditions and unpredictable device mobility, it becomes cumbersome to maintain a stable mmWave link for extended periods of time. Therefore, an alternative data dissemination mode considered here for the delivery of heavy multimedia data is to first cache it locally in the device and then, once the connection quality is restored, push the contents into the edge network infrastructure for further storage or immediate delivery to the intended recipient. This option is referred to as \textit{Store and Push} (see Fig.~\ref{fig:modes}(b)). 

Another viable alternative to compensate for the unpredictable fluctuations in mmWave connection quality is to rely on the opportunistic assistance of the `helper' IoT devices in proximity. By leveraging other moving machines within a fleet, the device with poor mmWave link may decide to forward its captured data to a neighboring device whenever the latter is sufficiently close. Such a behavior becomes beneficial whenever the proximate helper device has (or will soon have) a better quality link to the edge infrastructure. In this case, the helper device stores the forwarded data in its local cache and then pushes them into the edge network as soon as the channel quality is/becomes adequate. The use of collaborative in-device caching further lowers the data acquisition latency (by trading it for the increased utilization of in-device memory capacity) as well as improves the associated energy efficiency. Since the envisioned data dissemination mode involves forwarding of the contents to nearby device(s), it is called \textit{Forward and Push} hereinafter (see Fig.~\ref{fig:modes}(c)).

\begin{figure*}[!ht]
\centering
\includegraphics[width=1.6\columnwidth]{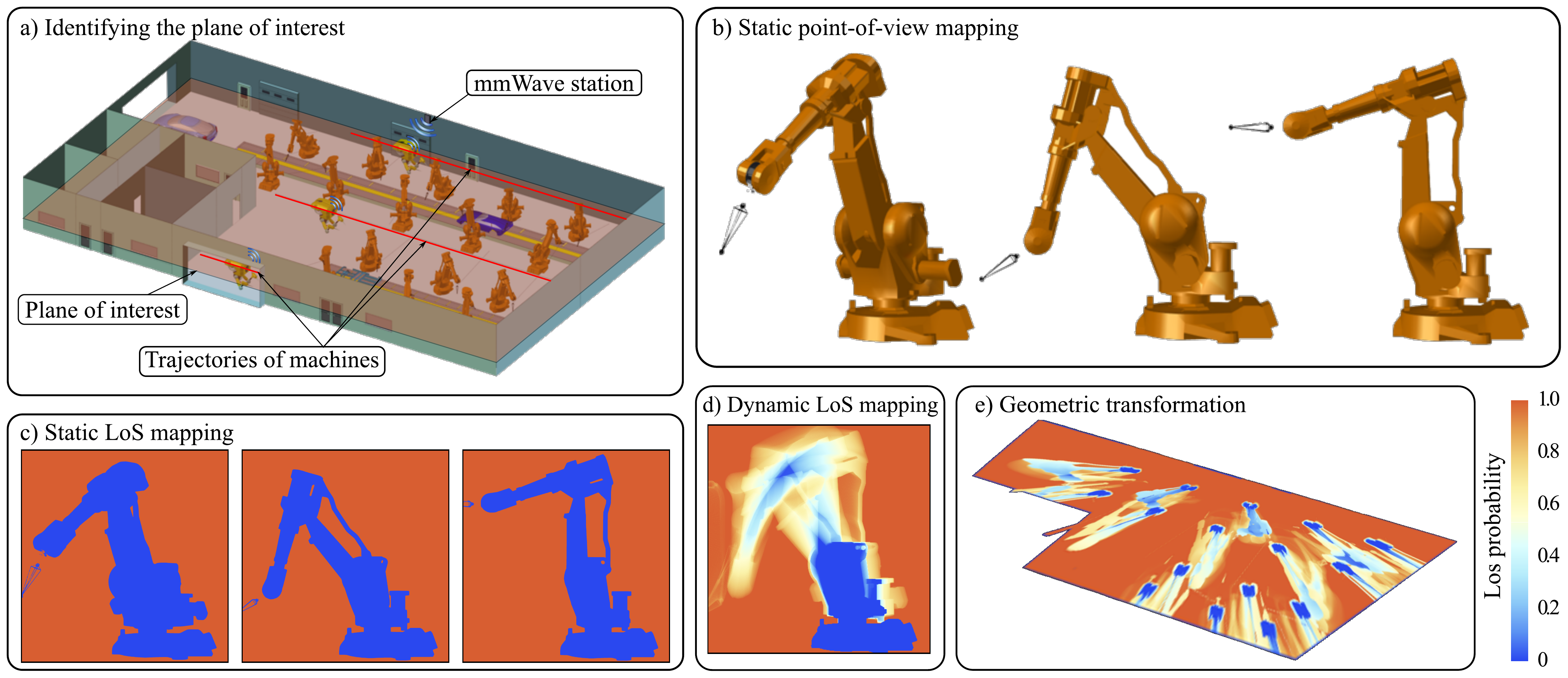}
\caption{Example application for the proposed LoS characterization methodology that captures moving patterns of industrial objects.}
\label{fig:methodology}
\vspace{-0.4cm}
\end{figure*}

As a result, the moving industrial machines generally have three distinct possibilities, which they may choose between intelligently. In doing so, they need to balance their experienced `infrastructure' link quality, the availability of caching helpers in proximity, and the properties of D2D channels toward them, among other factors. In contrast to any locally-optimal `hard' decisions as dictated by the preset contents delivery and placement strategies, the IoT devices may also leverage rich assistance information coming from the side of the network infrastructure. Such \textit{network-assisted} mode selection processes may be driven by `smooth' decision-making policies that involve system-wide knowledge of the immediate and \textit{expected} LoS conditions as well as employ appropriate device incentives. Most importantly, smart industrial machines may adapt their mode/relay selection strategies `on the fly' by applying information about the predicted residual time in the LoS/non-LoS state. Hence, the appropriate data dissemination mode may be selected \textit{proactively}, and a flexible dynamic methodology to employ such probabilistic knowledge is developed in the sequel.


\section{Predictive Mobility-Aware Channel Characterization}\label{sect:simulation}



In the use cases discussed above, there are essentially two main data transfer paths: (i) an \textit{infrastructure link}, where a moving device pushes its contents directly into the edge network via a serving mmWave station (which is stationary and may be deployed on a wall of the factory), and (ii) a \textit{D2D link}, where the device in question forwards the contents to a proximate IoT machine because of its poor infrastructure link conditions. Here, one of the key reasons for infrastructure link outage is dynamic blockage by moving objects in the scenario, which may occlude the LoS path. Therefore, we are primarily interested in characterizing the LoS probability as the key link quality indicator. This knowledge may then be utilized to design predictive mode selection policies. 


Given that the optimal mmWave station association rules are complicated for practical implementations~\cite{petrov2017dynamic}, we continue by sketching an example heuristic methodology that exploits the synergy between the mmWave radio channel knowledge and the available data dissemination modes. Going further, we also discuss its extensions to more complex and practical scenarios. Since factory automation environments are highly deterministic, we first employ our in-house 3D modeling software and then develop a suitable post-processing technique in what follows. We begin with characterizing the infrastructure link, where an industrial IoT device is served by a single mmWave station installed in a particular location. Hence, our approach needs to reconstruct a probabilistic LoS map that quantifies the chances of having an unobstructed mmWave path from an arbitrary point to the base station. 

A practical heuristic method is to employ photogrammetry techniques as displayed in Fig.~\ref{fig:methodology}. Accordingly, one first identifies a plane of interest as shown in Fig.~\ref{fig:methodology}(a). By means of 3D modeling, a set of images is obtained that capture all possible motions of the target IoT object inside the area with the desired granularity, see Fig.~\ref{fig:methodology}(b). All of the collected images are segmented by binarizing them based on the color of the plane considered in the first step, which creates raw LoS maps where the moving objects therein have a certain specific position (see Fig.~\ref{fig:methodology}(c)). The segmented images are then combined and the output is normalized as demonstrated in Fig.~\ref{fig:methodology}(d) and Fig.~\ref{fig:methodology}(e), respectively. The resultant image is then utilized to determine the LoS probability between a moving IoT device, which is located arbitrarily inside the area of interest, and its serving mmWave station.

Clearly, the characterization of D2D links calls for a more general approach, where the LoS probabilities are provided for two arbitrarily located moving devices due to \textit{dual} mobility. The corresponding process comprises two stages. First, the above set of preparatory steps needs to be completed for supplying the second stage with the probabilistic LoS data for each of the objects in question. As soon as the LoS blockage effects within the area of interest have been quantified, a similar procedure is applied at the second stage for every device type. The second stage of the proposed methodology delivers the LoS probability between two moving IoT devices located at their specified coordinates. The final probability of interest is a superposition of the LoS probabilities for each individual industrial object within the target factory area. The concluding step is to extract the sought LoS probability from the image that has been prepossessed in the first stage, and then repeat this procedure for every IoT object in the scenario.

Whenever a more detailed radio channel characterization is required or the trajectories of IoT devices are varying frequently with time, the described approach can be implemented on site. This can be realized by introducing a learning period, which is invoked every time when the device trajectory in question changes significantly. During this period, the IoT devices may collect information about their mmWave connection states with the serving base station as well as their neighbors. Once such state information is obtained, it is then propagated to the proximate IoT devices, while the entire system transitions to the operational state. Once subsequent alterations in the device trajectories are detected, the learning period is initiated again.

\begin{figure}[!ht]
\vspace{-0.2cm}
\centering
\includegraphics[width=0.9\columnwidth]{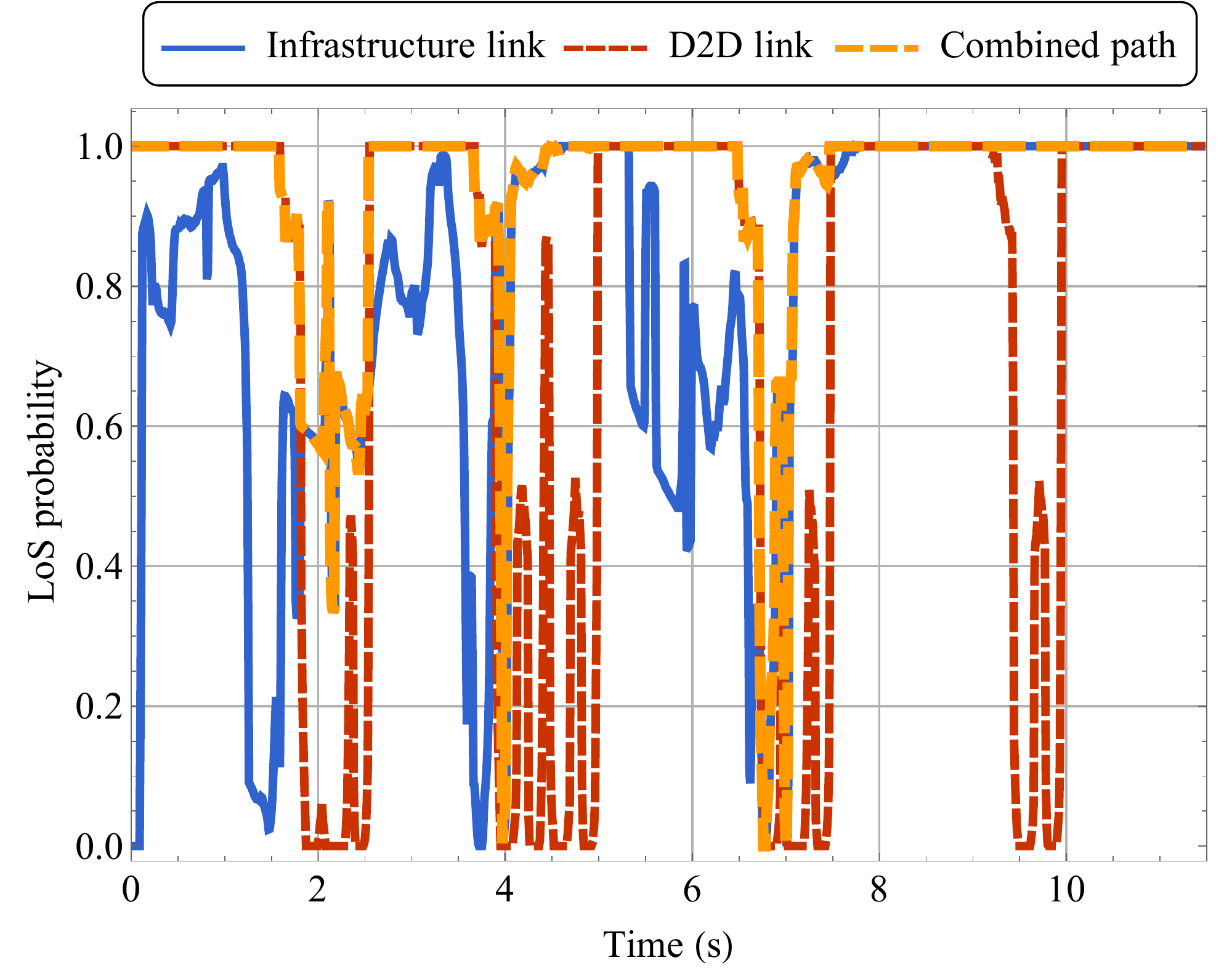}
\vspace{-0.2cm}
\caption{Example mobility-aware LoS dependence on time.}
\label{fig:LoS}
\vspace{-0.1cm}
\end{figure}

An illustration of the LoS blockage probabilities provided by the proposed methodology is shown in Fig.~\ref{fig:LoS} for three IoT machines, which move along the trajectories highlighted in Fig.~\ref{fig:methodology}(a). We observe that even in this simple scenario with all the trajectories having identical length, where IoT devices travel at the same constant speed, there is a high degree of spatial diversity on the mmWave links that can be exploited efficiently to improve the overall system-level performance. Particularly, at a certain point of time $t$, it is very probable to have at least one link to the serving station that remains entirely in the LoS conditions (hence offering the best choice of the data delivery path). 

Considering the data dissemination modes in Section~\ref{sec:up_schemes} and applying the system-wide knowledge of the LoS probability -- that varies with time subject to the actual moving patterns of different IoT objects -- it is then possible to construct and evaluate the data dissemination strategies, which are introduced and compared numerically in the following section. 


\section{Assessment of Caching-Aided Data Dissemination}

In this section, we report on the results of our simulation-based performance evaluation campaign that has been conducted to assess the data dissemination strategies. It delivers important numerical insights, which help understand the impact of collaborative D2D-enabled caching at the wireless edge. The parameters of interest are related to the proportion of data acquired before the completion deadline as well as the average delay in delivering the desired content. 


\subsection{Alternative Data Dissemination Strategies}

In our subsequent performance assessment, we consider three alternative data dissemination strategies that can be exploited concurrently by the moving industrial IoT devices in the target factory scenario to deliver their massive contents. The underlying objective is to disseminate data in a timely and reliable manner thus making the contents available to the factory operator(s) whenever requested. According to the data delivery modes introduced in Section~\ref{sec:up_schemes}, the first strategy considered here is \textit{Direct dissemination}. In this case, moving machines rely solely on the Direct Push transmission mode where data are sent to the edge network regardless of the current state of the infrastructure link. This simple approach does not take into account the potential performance degradation e.g., due to the mmWave link blockage by various obstacles as IoT devices move along their trajectories. 

Hence, an alternative solution is to consider a combined use of the Direct Push and the Store and Push dissemination modes by outlining \textit{Direct dissemination with storage} strategy. In this case, if a mobile IoT device experiences a significant decrease in the infrastructure link quality as predicted by utilizing the knowledge in Fig.~\ref{fig:LoS}, it caches the contents in its internal storage until when mmWave connectivity becomes available/feasible again. However, with this solution the contents delivery latency becomes a critical limiting factor, since the data may be kept in the internal memory for extended periods of time if the infrastructure link is unstable. 

To overcome this constraint, we introduce a smart data dissemination strategy named \textit{Predictive dissemination}, which intelligently leverages the spatial diversity across the available links (both infrastructure and D2D) at any given instant of time, see Fig.~\ref{fig:LoS}. It therefore exploits all three dissemination modes as per Section~\ref{sec:up_schemes}: in case the contents cannot be pushed into the edge network immediately (i.e., due to poor infrastructure connectivity), the mobile IoT device attempts to forward its data to the proximate D2D helpers, which will then be responsible for their delivery. Hence, caching the contents in the internal storage is only considered as the last-resort option, since it incurs additional latency in data delivery to the edge network. In Predictive dissemination, the target IoT device makes its mode selection decisions `on-the-fly' in accordance with the network-assistance information on the expected LoS conditions (see the methodology summarized in Section~\ref{sect:simulation} and the results provided in Fig.~\ref{fig:LoS}).

\subsection{Considered Factory Automation Deployment}

Our representative scenario of interest is a factory floor of [18x10]~m where 16 robots are deployed together with two conveyor belts. The restricted movements of these industrial machines are captured with the relevant mobility models that involve realistic rotations and tilts. The 5G mmWave radio technology is assumed to operate at 28~GHz with the bandwidth of 800~MHz. The target mmWave base station is deployed in the middle of the south wall inside the target area, see Fig. \ref{fig:methodology}(a). The radio channel conditions of the communicating machines are evaluated according to the methodology described in Section~\ref{sect:simulation} using the pathloss model proposed by 3GPP, which accounts for the necessary signal losses and fading effects. The D2D-specific neighbor discovery and connection establishment operations are managed directly by the 5G mmWave infrastructure as dictated by the ProSe functionality developed by 3GPP. The actual D2D links use the WiGig (802.11ad) protocol that has been standardized by the IEEE and operates at 60~GHz. 

We consider several mobile IoT devices that travel across the area of interest. Initially, they are distributed horizontally and at equal distance, while during simulations they move back and forth along the factory floor with the constant speed of 3 km/h. The moving machines disseminate their heavy traffic originating from the high-end sensors and cameras deployed on them as well as around them. Thus generated data need to be cached in the edge network before potential delivery to the human operator located in the remote control station. Contents are assumed to be undelivered (dropped) when the moving IoT device experiences a lack of connectivity or when the link budget is insufficient to complete the data transfer on time. The main modeling parameters are summarized in Table~\ref{tab:sim_parameters}.
  
\begin{table}[!ht]
\centering
\caption{Main modeling parameters}
\begin{tabular}{l|c}
\hline
\hline
\textbf{System parameter} & \textbf{Value}\\
\hline
\hline
mmWave carrier frequency & 28 GHz\\
Bandwidth of mmWave cellular & 800 MHz\\
WiGig carrier frequency & 60 GHz \\
Maximum WiGig radius & 100 m \\
mmWave station transmit power & 20 dBm \\
IoT device transmit power & 23 dBm \\ 
WiGig link setup time & 0.100 ms \\
WiGig target data rate & 10 Gbps \\
Number of simulation runs & 500 \\
\hline
\hline
\textbf{Application parameter} & \textbf{Value} \\
\hline
\hline
Video resolution & 4K (4096x4096), 120 FPS \\
Key-frame interval & [5-50]~ms \\
Max bit rate & 300 Mbps \\
Rate control & CBR \\
\hline
\hline
\end{tabular}
\label{tab:sim_parameters}
\vspace{-0.3cm}
\end{table}

\subsection{Representative Performance Results}\label{sec:results}


Our numerical assessment is conducted by relying on the network simulator 3 (ns-3) environment that is applied in conjunction with our in-house 3D modeling software~\cite{7592474}. In particular, we adopt the mmWave module developed by the New York University (NYU) team and publicly available on GitHub\footnote{ns-3 module for simulating mmWave-based cellular systems. Available at \url{https://github.com/nyuwireless/ns3-mmwave} [Accessed on 02/2018]}. The transmitted traffic is modeled after 4K VR 360$^{\circ}$ video streaming service with the resolution of 4096x4096 and the maximum bitrate of $300$ Mbps. The system-level metrics under consideration are: (i) the proportion of contents dropped due to dynamic LoS blockage, (ii) the proportion of contents dropped due to insufficient link data rate, and (iii) the average delay experienced in caching the contents inside the edge network. These parameters are evaluated by taking into account the intensity with which new contents are generated that corresponds to certain inter-arrival time of the video transfer requests.



\begin{figure}[!ht]
	\centering
	{\includegraphics[width=\columnwidth]{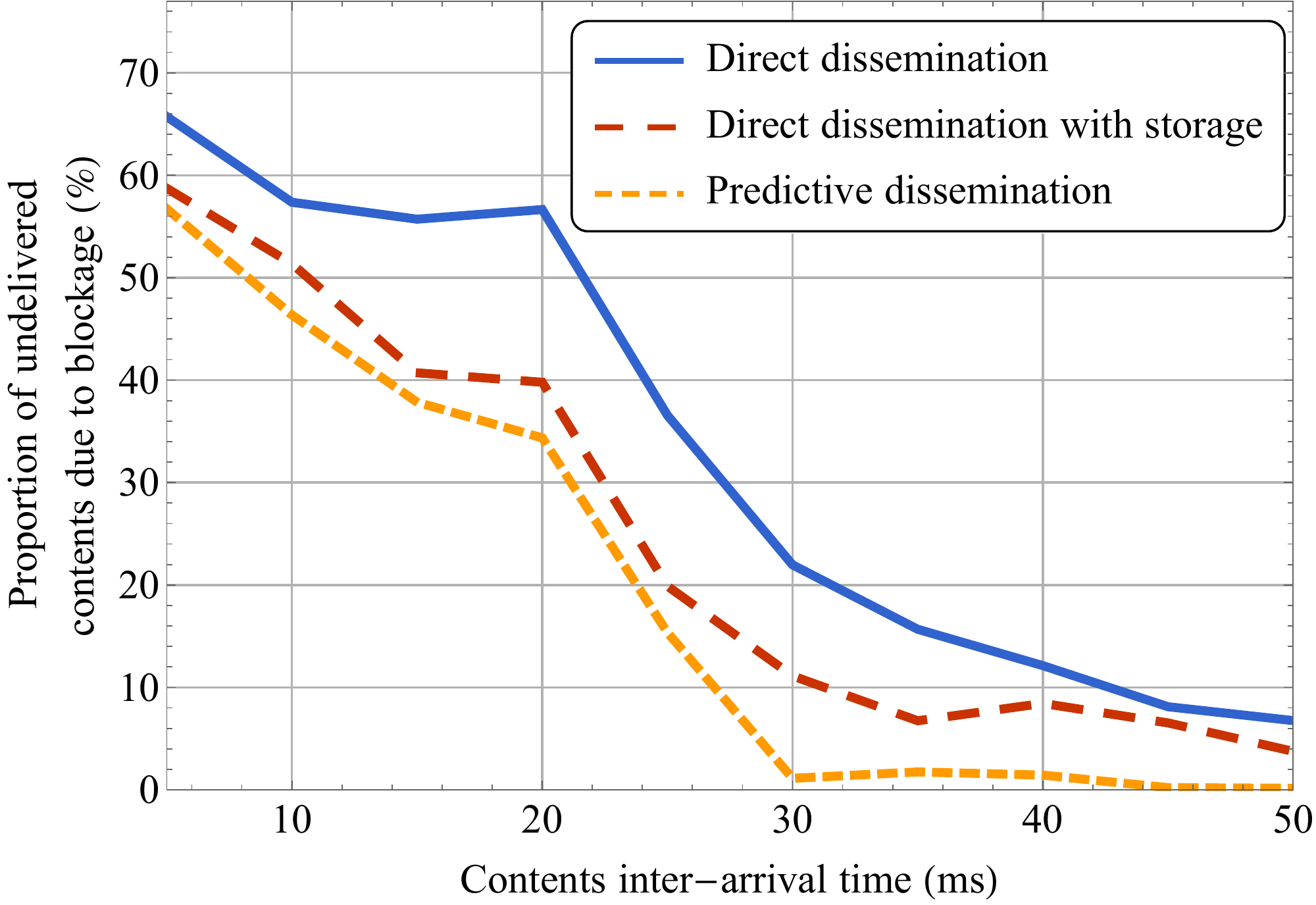}}	
	{\includegraphics[width=\columnwidth]{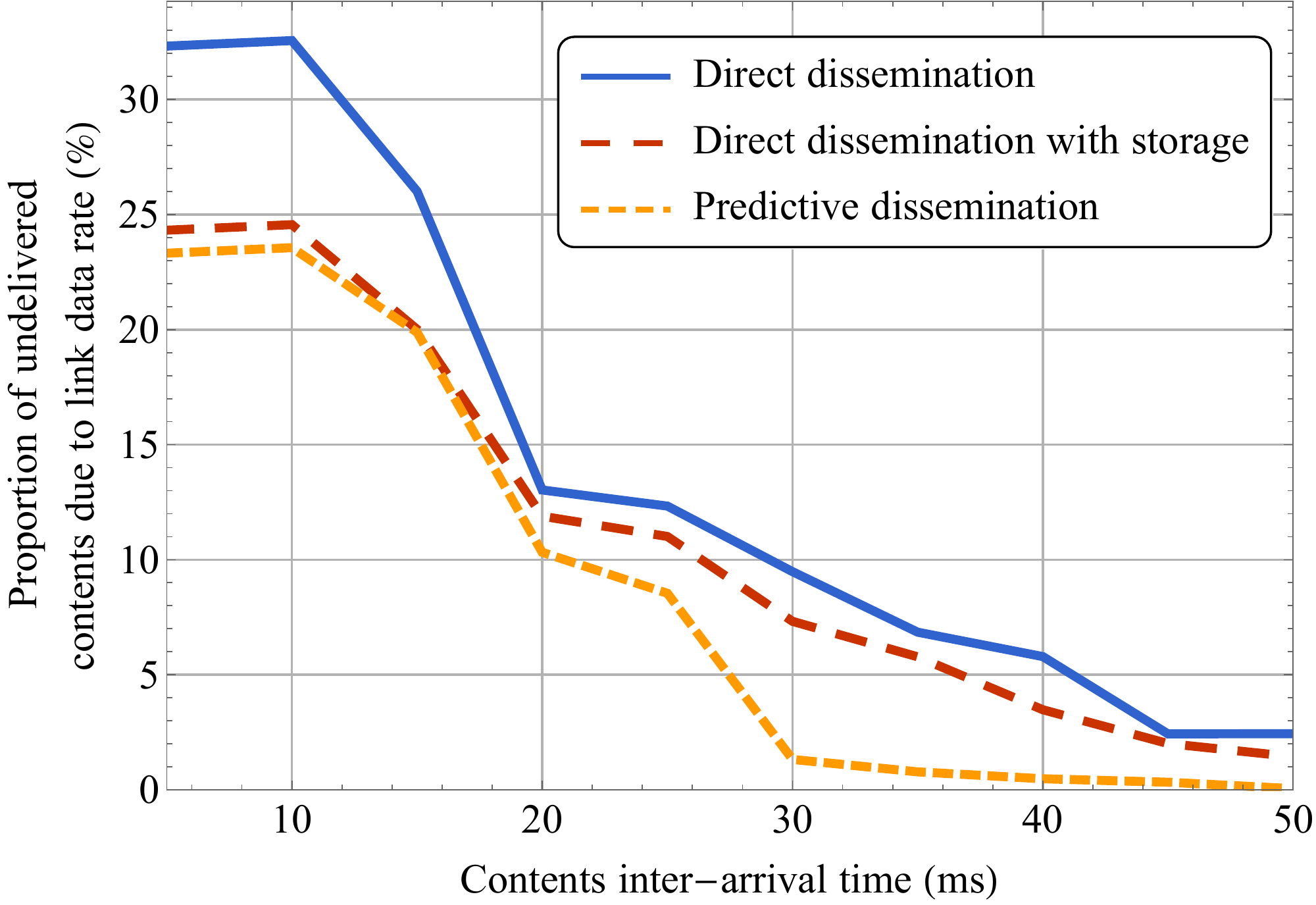}}	
	{\includegraphics[width=\columnwidth]{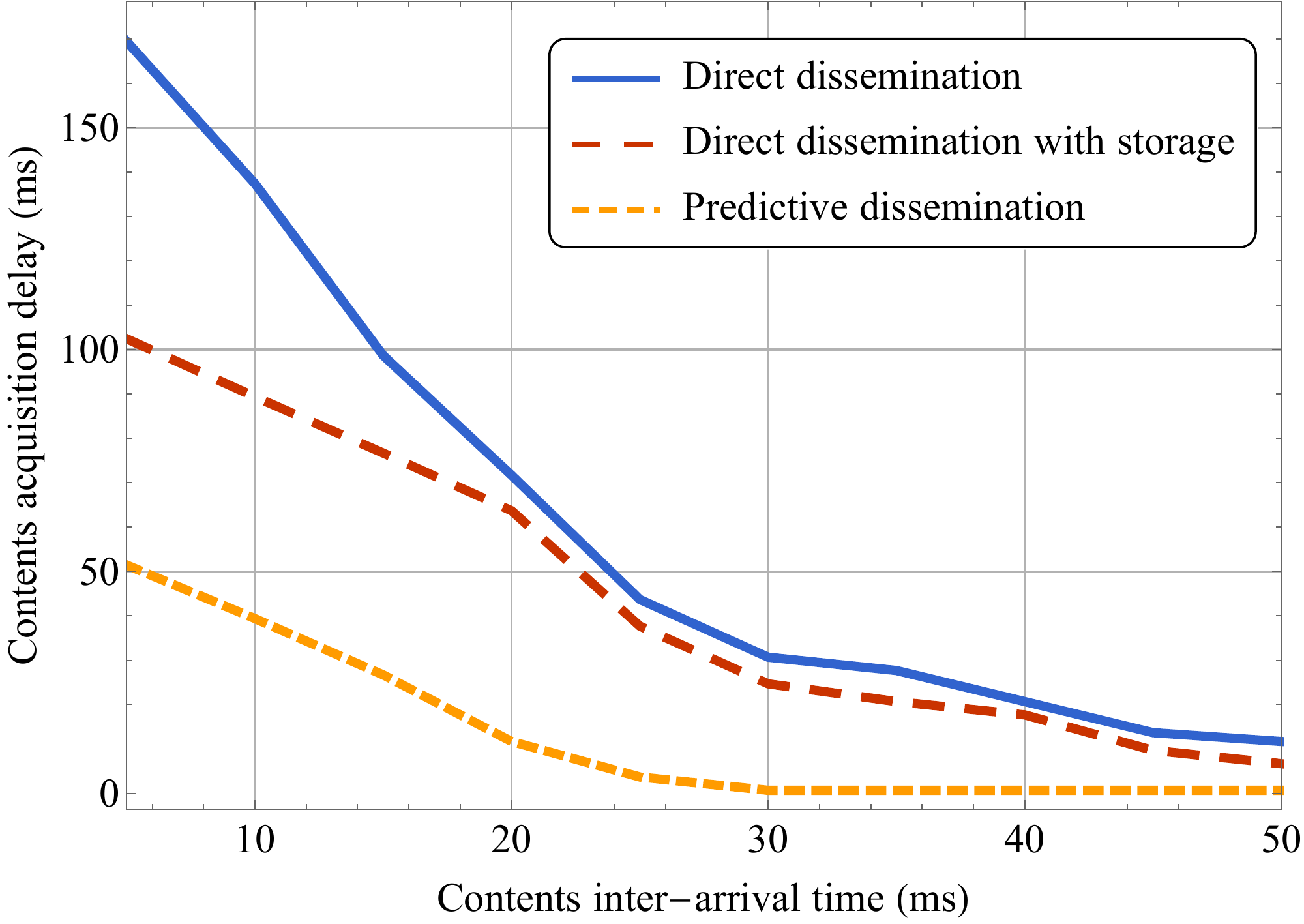}}
	\caption{Proportion of undelivered contents and average acquisition delay.}
	\label{fig:results}
	\vspace{-0.3cm}
\end{figure}

First, we focus on the proportion of undelivered contents that do not meet the target deadline because of mmWave propagation losses or blockage phenomena for different data arrival intensities. Fig.~\ref{fig:results}(top) and Fig.~\ref{fig:results}(middle) demonstrate both effects by varying the inter-arrival time from 5~ms to 50~ms. Recall that the top sub-figure refers to the case where the data do not reach the edge network timely due to LoS dynamics, whereas the middle sub-figure is related to the case where the actual link data rate is not sufficient to complete the acquisition. We learn from the figure that when the new requests are generated less frequently the chances of completing their acquisition on time increase. The \textit{Predictive dissemination} strategy shows the best performance as compared to its other two counterparts. 

Observe that the mmWave systems are generally able to support the data rates on the order of Gbps, thus requiring relatively little time to deliver the contents. However, as the arrival intensity grows, the IoT devices experience difficulties in transferring all of their heavy data. This situation is aggravated by the presence of factory equipment units deployed within our scenario of interest that may act as blockers for the mmWave radio signal. Another crucial conclusion is that forwarding the data to the neighboring devices via D2D links helps overcome these limitations and results in having more reliable connectivity. Further, it is also evident that obstacles are the dominant factor for the contents to be dropped in relation to the case where the link budget is insufficient. Indeed, this is an important learning that highlights the extent of performance degradation, which represents a serious challenge for the mmWave-based industrial system design.

Further, Fig.~\ref{fig:results}(bottom) demonstrates the average delay incurred when acquiring massive contents from the edge network storage. Here, the more intelligent \textit{Predictive dissemination} strategy outperforms the alternative solutions, since it exploits dynamic selection of the appropriate data transfer mode. In fact, this approach does not always forward the traffic to the closest IoT device in proximity, but adaptively selects the best delivery mode based on the LoS conditions. Accordingly, when the channel quality is adequate, the smarter predictive strategy pushes the data directly into the edge network over the infrastructure mmWave link. Otherwise, if there is no benefit in doing so, contents are forwarded via a D2D link to one of the helper devices, which are then responsible to cache the data and push them into the network as soon as the link condition becomes favorable. Naturally, if neither infrastructure link nor D2D link are available, the contents are stored in the originating IoT device and only pushed/forwarded whenever feasible.


\section{Main Conclusions}\label{sect:conclusions}

In this work, we introduced a novel methodology for the LoS path prediction that helps construct proactive data dissemination strategies in complex factory environments where industrial IoT devices travel along arbitrary trajectories and are surrounded by moving obstacles. Understanding that data transfer at mmWave frequencies is susceptible to sudden blockage and propagation losses, we therefore explored the benefits of collaborative in-device caching to improve delivery reliability and acquisition latency in mission-aware IoT operation. As a result, D2D-enabled caching at the wireless edge was shown to constitute a promising solution for augmenting system-level performance and the related practical challenges were discussed. 

In particular, we constructed and evaluated a predictive data dissemination strategy that exploits dynamic transmission mode selection where massive contents are pushed into the edge network immediately, forwarded to the proximate helper device for assistance, or cached inside the originating device until when the link recovers. With the proposed adaptive mode selection, our methodology delivers useful tools to enhance contents dissemination in realistic factory automation scenarios, mindful of device mobility and mmWave connection properties. Construction of pragmatic incentive-aware D2D relay and mode selection mechanisms for realistic industrial IoT setups becomes an attractive direction for further research.

\section*{Acknowledgment}

This work was supported by the Academy of Finland (projects WiFiUS and PRISMA) and by the project TAKE-5: The 5th Evolution Take of Wireless Communication Networks, funded by Tekes. The work of S. Andreev was supported in part by a Postdoctoral Researcher grant from the Academy of Finland and in part by a Jorma Ollila grant from Nokia Foundation.

\balance
\bibliographystyle{ieeetr}
\bibliography{paper}

\newpage

\section*{Authors' Biographies}

\textbf{Antonino Orsino} (antonino.orsino@gmail.com) is currently an experienced researcher at Ericsson Research, Finland. He received the B.Sc. degrees in Telecommunications Engineering from University Mediterranea of Reggio Calabria, Italy, in 2009 and the M.Sc. from University of Padova, Italy, in 2012. He also received his Ph.D. from University Mediterranea of Reggio Calabria, Italy, in 2017. He is actively working in 5G NR standardization activities and his current research interests include D2D and M2M communications in 4G/5G cellular systems, and IoT. He is the inventor/co-inventor of 10+ patent families, as well as the author/co-author of 50+ international scientific publications and numerous standardization contributions in the field of wireless networks.

\textbf{Roman Kovalchukov} (roman.kovalchukov@tut.fi) received the B.S. degree in fundamental informatics and information technologies from Peoples' Friendship University of Russia (RUDN University), Moscow, Russia, in July 2016. He is currently pursuing the M.S. degree in Information Technology at Tampere University of Technology, Tampere, Finland. He joined the Laboratory of Electronics and Communications Engineering of TUT as a research assistant in August 2016. His research interests are in wireless communications, with a focus on analysis of mmWave cellular networks, stochastic geometry, heterogeneous cellular networks, statistical modeling, and device-to-device cellular communications.

\textbf{Andrey Samuylov} (andrey.samuylov@tut.fi) received the Ms.C. in Applied Mathematics and Cand.Sc. in Physics and Mathematics from the RUDN University, Russia, in 2012 and 2015, respectively. Since 2015 he is working at Tampere University of Technology as a researcher, working on analytical performance analysis of various 5G wireless networks technologies. His research interests include P2P networks performance analysis, performance evaluation of wireless networks with enabled D2D communications, and mmWave-band communications.

\textbf{Dmitri Moltchanov} (dmitri.moltchanov@tut.fi) received the M.Sc. and Cand.Sc. degrees from the St. Petersburg State University of Telecommunications, Russia, in 2000 and 2002, respectively, and the Ph.D. degree from the Tampere University of Technology in 2006. He is a senior research scientist with the Laboratory of Electronics and Communications Engineering, Tampere University of Technology, Finland. He has authored over 100 publications. His research interests include performance evaluation and optimization issues of wired and wireless IP networks, Internet traffic dynamics, quality of user experience of real-time applications, and mmWave/terahertz communications systems. He serves as a TPC member in a number of international conferences.

\textbf{Sergey Andreev} (sergey.andreev@tut.fi) is a senior research scientist in the Laboratory of Electronics and Communications Engineering at Tampere University of Technology, Finland. He received the Specialist degree (2006) and the Cand.Sc. degree (2009) both from St. Petersburg State University of Aerospace Instrumentation, St. Petersburg, Russia, as well as the Ph.D. degree (2012) from Tampere University of Technology. Sergey (co-)authored more than 100 published research works on wireless communications, energy efficiency, heterogeneous networking, cooperative communications, and machine-to-machine applications.

\textbf{Yevgeni Koucheryavy} (evgeni.kucheryavy@tut.fi) is a full professor at the Laboratory of Electronics and Communications Engineering of Tampere University of Technology (TUT), Finland. He received his Ph.D. degree (2004) from TUT. He is the author of numerous publications in the field of advanced wired and wireless networking and communications. He is Associate Technical Editor of IEEE Communications Magazine and Editor of IEEE Communications Surveys and Tutorials.

\textbf{Mikko Valkama} (mikko.e.valkama@tut.fi) received his M.Sc. and D.Sc. degrees (both with honors) from Tampere University of Technology, Finland, in 2000 and 2001, respectively. In 2003, he worked as a visiting researcher at San Diego State University, California. Currently, he is a full professor and Head of the Laboratory of Electronics and Communications Engineering at Tampere University of Technology. His research interests include radio communications and systems with particular emphasis on 5G and beyond mobile communications.

\end{document}